\documentclass[a4paper,12pt]{article}

\usepackage{ifpdf}

\newif\ifpdf
\ifx\pdfoutput\undefined
  \pdffalse
\else
  \pdfoutput=1
  \pdftrue
\fi

\RequirePackage{xspace} %
\RequirePackage{subfigure} %
\RequirePackage[centertags]{amsmath} %
\RequirePackage{amssymb}
\RequirePackage{wrapfig} %
\RequirePackage{calc} %
\RequirePackage{ifthen}
\RequirePackage{tabularx} %
\RequirePackage{flafter} %
\RequirePackage{fancyhdr} %

\ifpdf
  \RequirePackage[pdftex]{color}%
  \RequirePackage{colortbl}%
  \RequirePackage{array}%
  \RequirePackage[pdftex]{graphicx}

  \RequirePackage[ pdftex, plainpages = false, pdfpagelabels,
                 pdfpagelayout = useoutlines,
                 bookmarks,
                 breaklinks = true,
                 linktocpage,
                 pagebackref,                      
                 colorlinks = true,
                 linkcolor = blue,
                 urlcolor  = blue,
                 citecolor = blue,
                 anchorcolor = blue,
                 hyperindex = true,
                 hyperfigures
                 ]{hyperref}

\else
  \RequirePackage{color}
  \RequirePackage{colortbl}
   \RequirePackage{array}
  \RequirePackage[dvips]{graphicx}
  \RequirePackage{hyperref}
  \usepackage{rotating}
\fi

\usepackage{makeidx} 
\usepackage{setspace} 
\usepackage{rotating} 
\usepackage{ecltree}
\usepackage{epic}
\usepackage{supertabular}  
\usepackage{color}
\usepackage{exscale}
\usepackage{fontenc}
\usepackage{ifthen}
\usepackage{latexsym}
\usepackage{makeidx}
\usepackage{syntonly}
\usepackage{inputenc}
\usepackage{graphicx}
\usepackage{setspace}
\usepackage{caption2}
\usepackage[english]{babel}
\usepackage[square, comma,numbers,sort&compress]{natbib}
\usepackage{hypernat}
\usepackage{boxedminipage}
\usepackage{framed}
\usepackage{longtable}
\usepackage[all]{hypcap}    
\usepackage{algorithm2e}
\usepackage{algorithmic}
\usepackage{lscape}
\usepackage{pdflscape}
\usepackage[T1]{fontenc}
\usepackage{microtype}

\newcommand{\note}[1]{\marginpar[left]{\singlespace \tiny #1}}

\renewcommand{\sectionmark}[1]%
      {\markright{\thesection\ #1}} 

\renewcommand{\note}[1]{}

\newcommand{\CIF}     {\centering \includegraphics[width=2.7in]} %
\newcommand{\Hs}      {\hspace{-0.5cm}} %

\doublespace

\DisableLigatures[f]{encoding = *, family = * }

\begin{document}
\begin{center}
{\Large Analytical solutions for the flow of Carreau and Cross fluids in circular pipes and thin
slits}
\par\end{center}{\Large \par}

\begin{center}
Taha Sochi
\par\end{center}

\begin{center}
{\scriptsize University College London, Department of Physics \& Astronomy, Gower Street, London,
WC1E 6BT \\ Email: t.sochi@ucl.ac.uk.}
\par\end{center}

\begin{abstract}
\noindent
In this paper, analytical expressions correlating the volumetric flow rate to the pressure drop are
derived for the flow of Carreau and Cross fluids through straight rigid circular uniform pipes and
long thin slits. The derivation is based on the application of
Weissenberg-Rabinowitsch-Mooney-Schofield method to obtain flow solutions for generalized Newtonian
fluids through pipes and our adaptation of this method to the flow through slits. The derived
expressions are validated by comparing their solutions to the solutions obtained from direct
numerical integration. They are also validated by comparison to the solutions obtained from the
variational method which we proposed previously. In all the investigated cases, the three methods
agree very well. The agreement with the variational method also lends more support to this method
and to the variational principle which the method is based upon. \vspace{0.3cm}


\par\end{abstract}

\begin{center}

\par\end{center}

\section{Introduction} \label{Introduction}

There are many fluid models that have been developed and employed in the recent decades to describe
and predict the bulk and {\it in situ} rheology of non-Newtonian fluids. Amongst these, Carreau,
and to a certain degree Cross, are distinguished by their popularity and widespread use especially
in modeling the rheological behavior of biological fluids and polymeric liquids. Both are
four-parameter models that depend on the low-shear and high-shear viscosities, a characteristic
time or strain rate constant and a flow behavior exponent. They are usually used to describe the
time-independent shear-thinning category of the non-Newtonian fluids.

These widely used models provide good match to the experimental data in many flow situations, such
as the flow in the blood arteries and through porous media as well as rheometric measurements, and
hence they are popular in various biological, technological and industrial disciplines, such as
biosciences and engineering, reservoir engineering and food processing. For example they are
systematically used, in their different variants and forms and their modified versions, to model
and predict the flow of biological fluids like blood, and polymeric liquids like Xanthan gum and
polyacrylamide gel solutions \cite{CannellaHS1988, SorbieCJ1989, GijsenVJ1998, GijsenAVJ1999,
Georgiou2003, LopezVB2003, LopezB2004, AbrahamBH2005, BoxGRR2005, ChenLW2006, PerrinTSC2006,
JonasovaV2008, KimVL2008, VimmrJ2008, LukacovaZ2008, FisherR2009, SankarI2009, ZauskovaM2010,
LiuT2011, MollaP2012, DesplanquesRGM2012, ToscoMLS2013}.

To the best of our knowledge, no analytical solutions that correlate the volumetric flow rate to
the pressure drop in confined geometries, specifically pipes and slits, have been reported in the
literature for these fluids \cite{LopezVB2003, Lopezthesis2004, BalhoffRKMP2012, CastroOSB2014}
despite their wide application. This is mainly due to the rather complicated expressions that
result from applying the traditional methods of fluid mechanics to these models in any analytical
derivation treatment. Therefore, the users of these models either employ empirical approximations
or use numerical approaches which normally utilize mesh-based techniques like finite element and
finite difference methods.

In this paper, we make an attempt to derive fully analytical solutions for the flow of these fluids
through straight rigid circular uniform tubes and thin long slits. We use a method attributed to
Weissenberg, Rabinowitsch, Mooney and Schofield \cite{Skellandbook1967}, and may be others, and
hence we call it WRMS method. The method, as reported in the literature, is customized to the flow
in circular pipes; and hence we adapt it to the flow in thin slits to obtain flow relations for
this type of conduit geometry as well.

In fact there are two objectives to the present paper. The first is the derivation of the
analytical expressions, as outlined in the previous paragraph, which is useful, and may even be
necessary, in various rheological and fluid mechanical applications. The second, which is not less
important, is the solidification and support to our recently proposed \cite{SochiVariational2013,
SochiSlitPaper2014, SochiVarNonNewt2014} variational method which is based on optimizing the total
stress in the flow conduit by applying the Euler-Lagrange principle to find totally or partly
analytical flow relations for generalized Newtonian fluids through various conduit geometries. As
we will see, the results of the newly derived formulae in the present paper for the flow of Carreau
and Cross fluids through pipes and slits agree very well with the results obtained from the
variational method. Since the two methods are totally independent and are based on completely
different theoretical and mathematical infrastructures, they provide support and validation to each
other.

The plan for this paper is that in section \S\ \ref{Method} we present the WRMS method for the flow
of generalized Newtonian fluids through pipes and derive its adaptation for the flow through slits.
The WRMS is then applied in section \S\ \ref{Pipe} to derive analytical relations for the flow of
Carreau and Cross fluids through pipes, while its adaptation is used in section \S\ \ref{Slit} to
derive these relations for the flow through slits. The derived analytical expressions are then
validated in section \S\ \ref{Validation} by numerical integration and by comparison to the flow
solutions which are obtained from the variational method. The paper is ended in section \S\
\ref{Conclusions} with short discussion and conclusions about the purpose and the achieved
objectives of this study and possible future extensions to other models.

\section{Method}\label{Method}

First, we should state our assumptions about the flow, fluid and geometry and mechanical properties
of the employed conduits. In this investigation we assume a laminar, incompressible, isothermal,
steady, pressure-driven, fully-developed flow of a time-independent, purely-viscous fluid that is
properly described by the generalized Newtonian fluid model where the viscosity depends only on the
contemporary rate of strain and hence it has no deformation-dependent memory. As indicated earlier,
the pipe is assumed to be straight rigid with a uniform and circularly shaped cross sectional area
while the slit is assumed to be straight rigid long and thin with a uniform cross section. It is
also presumed that the entry and exit edge effects and external body forces, like gravity, are
negligible. As for the boundary conditions, we assume no-slip at the tube and slit walls
\citep{SochiSlip2011} with the flow velocity profile having a stationary derivative point at the
symmetry center line of the tube and symmetry center plane of the slit which means that the profile
has a blunt rounded vertex.

\begin{figure}[!h]
\centering{}
\includegraphics
[scale=0.65] {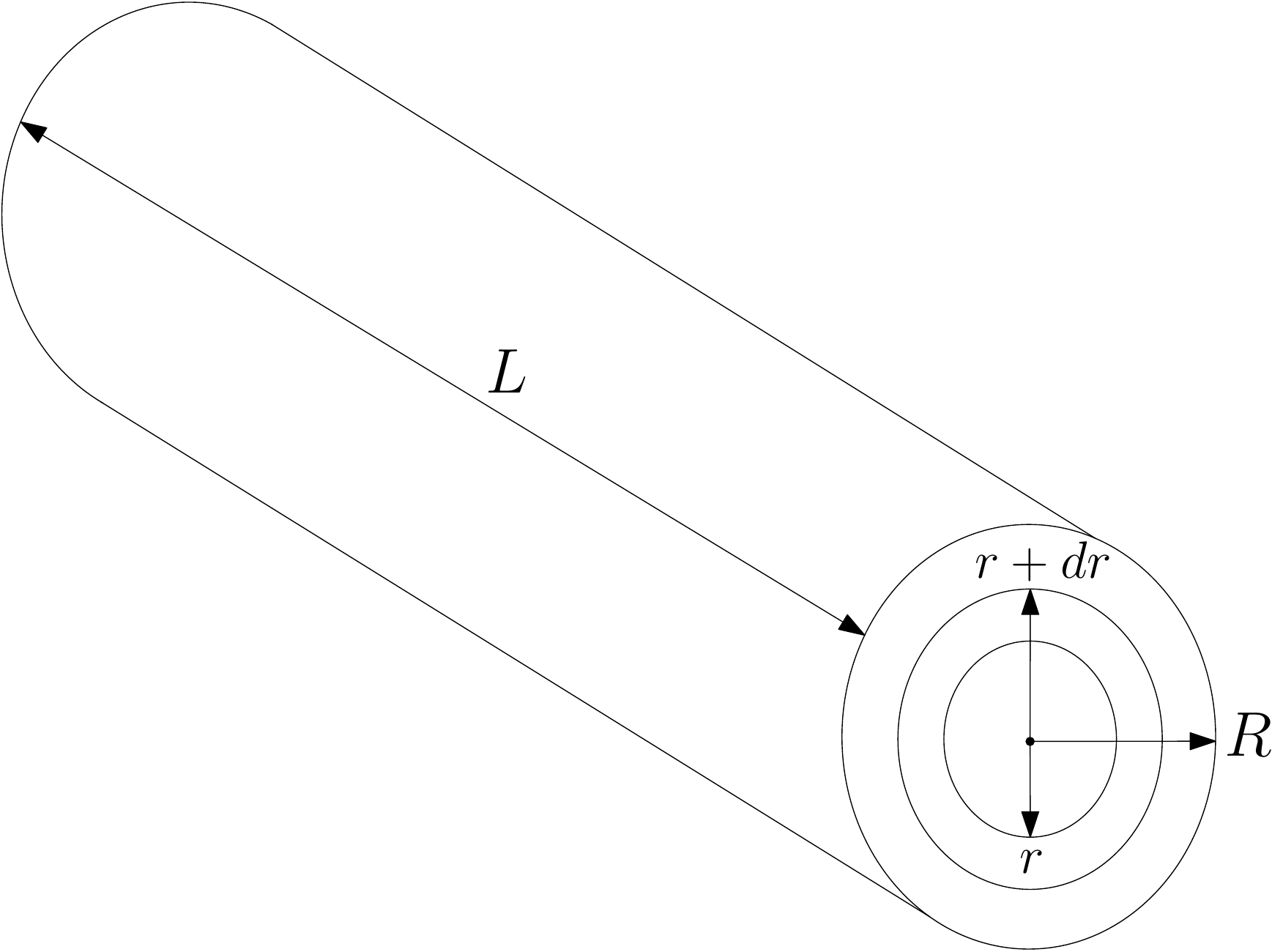} \caption{A diagram depicting the geometry of the straight rigid circular
uniform pipe that is used in the current study.} \label{PipePlot}
\end{figure}

We first present the derivation of the general formula for the volumetric flow rate of generalized
Newtonian fluids through pipes that satisfy the above-stated assumptions, as depicted in Figure
\ref{PipePlot}. The derivation is attributed to Weissenberg, Rabinowitsch, Mooney, and Schofield
and hence we label it with WRMS. The outline of this derivation is a modified version of what is in
\cite{Skellandbook1967} which we reproduce here for the purpose of availability, clarity and
completeness. We then follow this by adapting the WRMS method from the pipe geometry to the long
thin slit geometry which we also call it WRMS method. The difference between the two will be
obvious from the context.

The differential volumetric flow rate in a differential annulus between $r$ and $r+dr$, as depicted
in Figure \ref{PipePlot}, is given by

\begin{equation}
dQ=v2\pi rdr
\end{equation}
where $Q$ is the volumetric flow rate, $r$ is the radius and $v\equiv v(r)$ is the fluid velocity
at $r$ in the axial direction. The total flow rate is then obtained from integrating the
differential flow rate between $r=0$ and $r=R$ where $R$ is the tube radius, i.e.

\begin{equation}
Q=\pi\int_{0}^{R}v2rdr=\pi\int_{0}^{R^{2}}vd(r^{2})
\end{equation}
On integrating by parts we obtain

\begin{equation}
Q=\pi\left[vr^{2}-\int r^{2}dv\right]_{0}^{R^{2}}\label{eqQ1}
\end{equation}
The first term on the right hand side of the last equation is zero because at the lower limit
$r=0$, and at the upper limit $v=0$ due to the no-slip boundary condition \cite{SochiSlip2011};
moreover

\begin{equation}
dv=-\gamma_r dr
\end{equation}
where $\gamma_r\equiv\gamma(r)$ is the rate of shear strain at $r$. On eliminating the first term,
substituting for $dv$ and changing the limits of integration, Equation \ref{eqQ1} becomes

\begin{equation}
Q=\pi\int_{0}^{R}r^{2}\gamma_r dr\label{eqQ2}
\end{equation}
Now, by definition, the shear stress at the tube wall, $\tau_{w}$, is the ratio of the force normal
to the tube cross section, $F_{\perp}$, to the area of the internal surface parallel to this force,
$A_{\parallel}$, that is

\begin{equation}\label{tauw}
\tau_{w}\equiv\tau(r=R)\equiv\tau_{R}=\frac{F_{\perp}}{A_{\parallel}}=\frac{\pi R^{2}\Delta p}{2\pi
RL}=\frac{R\,\Delta p}{2L}
\end{equation}
where $L$ is the tube length, and $\Delta p$ is the pressure drop across the tube. Similarly

\begin{equation}
\tau(r)\equiv\tau_{r}=\frac{r\Delta p}{2L}\,\,\,\,\,\,\,\,\,\,\,\,\,\,\,\,\,\,\,\,(0\le
r<R)
\end{equation}
Hence

\begin{equation}
\frac{\tau_{r}}{\tau_{R}}=\frac{r}{R}\,\,\,\,\,\,\,\,\,\,\rightarrow\,\,\,\,\,\,\,\,\,
r^{2}=\frac{R^{2}\tau_{r}^{2}}{\tau_{R}^{2}}
\end{equation}
and

\begin{equation}
\frac{dr}{d\tau_{r}}=\frac{R}{\tau_{R}}\,\,\,\,\,\,\,\,\,\,\,\,\rightarrow\,\,\,\,\,\,\,\,\,\,\,
dr=\frac{R}{\tau_{R}}d\tau_{r}
\end{equation}
On substituting $r^{2}$ and $dr$ from the last two equations into Equation \ref{eqQ2} and changing
the limits of integration we get

\begin{equation}
Q=\pi\int_{0}^{\tau_{R}}\frac{R^{2}\tau_{r}^{2}}{\tau_{R}^{2}}\gamma_r\frac{R}{\tau_{R}}d\tau_{r}=\frac{\pi
R^{3}}{\tau_{R}^{3}}\int_{0}^{\tau_{R}}\tau_{r}^{2}\gamma_r d\tau_{r}
\end{equation}
that is

\begin{equation}
\boxed{ Q=\frac{\pi R^{3}}{\tau_{w}^{3}}\int_{0}^{\tau_{w}}\gamma \tau^{2} d\tau }
\end{equation}
where it is understood that $\gamma=\gamma_{r}\equiv\gamma(r)$ and $\tau=\tau_{r}\equiv\tau(r)$.


\begin{figure}[!t]
\centering{}
\includegraphics
[scale=0.75] {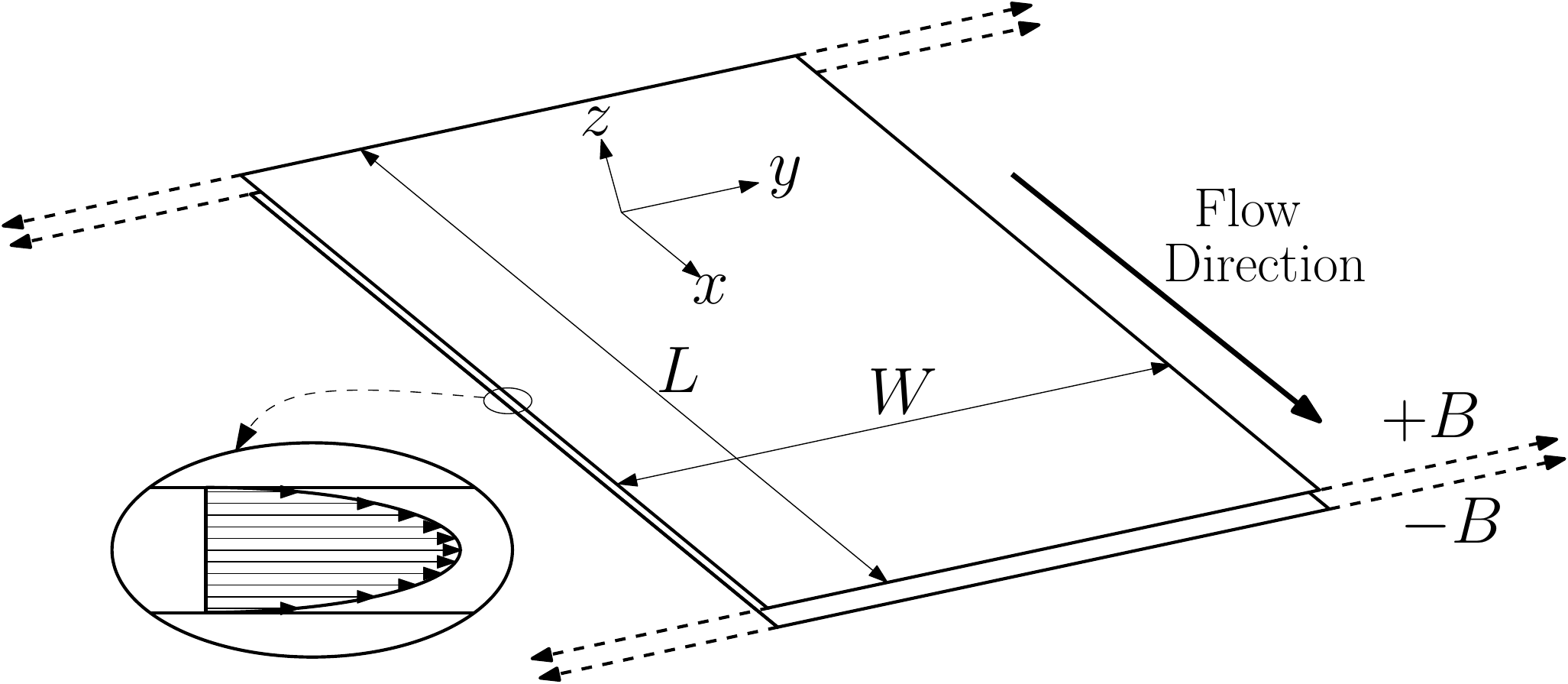} \caption{A diagram depicting the geometry of the straight rigid uniform
long thin slit that is used in the current study.} \label{SlitPlot}
\end{figure}

Next, we derive a general formula for the volumetric flow rate of generalized Newtonian fluids
through a long thin slit, depicted in Figure \ref{SlitPlot}, by adapting the WRMS method, which is
described in the last paragraph, as follow. The differential flow rate through a differential strip
along the slit width is given by

\begin{equation}
dQ=vWdz
\end{equation}
where $W$ is the slit width and $v\equiv v(z)$ is the fluid velocity at $z$ in the $x$ direction
according to the coordinates system demonstrated in Figure \ref{SlitPlot}. Hence

\begin{equation}
\frac{Q}{W}=\int_{-B}^{+B}vdz
\end{equation}
where $B$ is the slit half height with the slit being positioned symmetrically with respect to the
plane $z=0$. On integrating by parts we get

\begin{equation}
\frac{Q}{W}=\left[vz\right]_{-B}^{+B}-\int_{v_{-B}}^{v_{+B}}zdv
\end{equation}
The first term on the right hand side of the last equation is zero due to the no-slip boundary
condition at the slit walls, and hence we have

\begin{equation}
\frac{Q}{W}=-\int_{v_{-B}}^{v_{+B}}zdv\label{eqQ}
\end{equation}
Now, if we follow a similar argument to that of pipe, the shear stress at the slit walls,
$\tau_{w}$, will be given by

\begin{equation}
\tau_{w}\equiv\tau_{\pm B}=\frac{F_{\perp}}{A_{\parallel}}=\frac{2BW\Delta p}{2WL}=\frac{B \Delta
p}{L}
\end{equation}
where $L$ is the slit length and $\Delta p$ is the pressure drop across the slit. Similarly we have

\begin{equation}
\tau_{z}=\frac{z\Delta p}{L}
\end{equation}
where $\tau_{z}$ is the shear stress at $z$. Hence

\begin{equation}
\tau_{z}=\frac{z}{B}\tau_{\pm B}\,\,\,\,\,\,\,\,\rightarrow\,\,\,\,\,\,\,\,\,
z=\frac{B\tau_{z}}{\tau_{\pm B}}\,\,\,\,\,\,\,\,\rightarrow\,\,\,\,\,\,\,\,\,
dz=\frac{Bd\tau_{z}}{\tau_{\pm B}}\label{eqAAA}
\end{equation}
We also have

\begin{equation}
\gamma_z=-\frac{dv}{dz}\,\,\,\,\,\,\,\,\rightarrow\,\,\,\,\,\,\,\,\, dv=-\gamma_z
dz=-\gamma_z\frac{Bd\tau_{z}}{\tau_{\pm B}}\label{eqBBB}
\end{equation}
Now due to the symmetry with respect to the plane $z=0$ we have

\begin{equation}
\tau_{B}\equiv\tau_{+B}=\tau_{-B}
\end{equation}
On substituting from Equations \ref{eqAAA} and \ref{eqBBB} into Equation \ref{eqQ}, considering the
symmetry with respect to the center plane $z=0$ and changing the limits of integration we obtain

\begin{equation}
\frac{Q}{W}=\int_{\tau_{-B}}^{\tau_{+B}}\frac{B\tau_{z}}{\tau_{\pm
B}}\gamma_z\frac{Bd\tau_{z}}{\tau_{\pm
B}}=2\left(\frac{B}{\tau_{B}}\right)^{2}\int_{0}^{\tau_{B}}\tau_{z}\gamma_z d\tau_{z}
\end{equation}
that is

\begin{equation}
\boxed{ Q=2W\left(\frac{B}{\tau_{w}}\right)^{2}\int_{0}^{\tau_{w}}\gamma\tau d\tau }
\end{equation}
where it is understood that $\gamma=\gamma_{z}\equiv\gamma(z)$ and $\tau=\tau_{z}\equiv\tau(z)$.

\section{Pipe Flow}\label{Pipe}

In the following two subsections we apply the WRMS method to derive analytical expressions for the
flow of Carreau and Cross fluids in straight rigid circular uniform pipes.

\subsection{Carreau}

For Carreau fluids, the viscosity is given by \cite{Sorbiebook1991, Tannerbook2000}

\begin{equation}
\mu=\frac{\tau}{\gamma}=\mu_{i}+\left(\mu_{o}-\mu_{i}\right)\left(1+\lambda^{2}\gamma^{2}\right)^{\left(n-1\right)/2}
\end{equation}
where $\mu_{o}$ is the low-shear viscosity, $\mu_{i}$ is the high-shear viscosity, $\lambda$ is a
characteristic time constant, and $n$ is the flow behavior index. This, for the sake of
compactness, can be written as

\begin{equation}\label{eqCarVis}
\mu=\frac{\tau}{\gamma}=\mu_{i}+\delta\left(1+\lambda^{2}\gamma^{2}\right)^{n'/2}
\end{equation}
where $\delta=\left(\mu_{o}-\mu_{i}\right)$, and $n'=\left(n-1\right)$. Therefore

\begin{equation}
\tau=\gamma\left[\mu_{i}+\delta\left(1+\lambda^{2}\gamma^{2}\right)^{n'/2}\right]\label{tauEqCar}
\end{equation}
From WRMS method we have

\begin{equation}
\frac{Q\,\tau_{w}^{3}}{\pi R^{3}}=\int_{0}^{\tau_{w}}\gamma\tau^{2} d\tau\label{eqWRMSTu}
\end{equation}
If we label the integral on the right hand side of Equation \ref{eqWRMSTu} with $I$ and substitute
$\tau$ from Equation \ref{tauEqCar} into $I$ we obtain

\begin{equation}
I=\int_{0}^{\tau_{w}}\gamma^{3}\left[\mu_{i}+\delta\left(1+\lambda^{2}\gamma^{2}\right)^{n'/2}\right]^{2}d\tau\label{eqICarTu}
\end{equation}
Now, from Equation \ref{tauEqCar} we have

\begin{equation}
d\tau=\left[\mu_{i}+\delta\left(1+\lambda^{2}\gamma^{2}\right)^{n'/2}+n'\delta\lambda^{2}\gamma^{2}\left(1+\lambda^{2}\gamma^{2}\right)^{(n'-2)/2}\right]d\gamma\label{eqdtauCar}
\end{equation}
On substituting from Equation \ref{eqdtauCar} into Equation \ref{eqICarTu} and changing the
integration limits we get

\begin{equation}
I=\int_{0}^{\gamma_{w}}\gamma^{3}\left[\mu_{i}+\delta\left(1+\lambda^{2}\gamma^{2}\right)^{n'/2}\right]^{2}\left[\mu_{i}+\delta\left(1+\lambda^{2}\gamma^{2}\right)^{n'/2}+n'\delta\lambda^{2}\gamma^{2}\left(1+\lambda^{2}\gamma^{2}\right)^{(n'-2)/2}\right]d\gamma\label{eqICarTu2}
\end{equation}
where $\gamma_{w}$ is the rate of shear strain at the tube wall. On solving this integral equation
analytically and evaluating it at its two limits we obtain

\begin{eqnarray}\label{eqICarTu3}
I & = & \frac{\delta^{3}\left[3\lambda^{4}\left(3n'^{2}+5n'+2\right)\gamma_{w}^{4}-3n'\lambda^{2}\gamma_{w}^{2}+2\right]\left(1+\lambda^{2}\gamma_{w}^{2}\right)^{3n'/2}}{3\lambda^{4}\left(9n'^{2}+18n'+8\right)} \nonumber \\
 & + & \frac{\mu_{i}\delta^{2}\left[\lambda^{4}\left(2n'^{2}+5n'+3\right)\gamma_{w}^{4}-n'\lambda^{2}\gamma_{w}^{2}+1\right]\left(1+\lambda^{2}\gamma_{w}^{2}\right)^{n'}}{2\lambda^{4}\left(n'+1\right)\left(n'+2\right)} \nonumber \\
 & + & \frac{\mu_{i}^{2}\delta\left[\lambda^{4}\left(n'^{2}+5n'+6\right)\gamma_{w}^{4}-n'\lambda^{2}\gamma_{w}^{2}+2\right]\left(1+\lambda^{2}\gamma_{w}^{2}\right)^{n'/2}}{\lambda^{4}\left(n'+2\right)\left(n'+4\right)}+\frac{\mu_{i}^{3}\gamma_{w}^{4}}{4} \nonumber \\
 & - & \left(\frac{2\delta^{3}}{3\lambda^{4}\left(9n'^{2}+18n'+8\right)}+\frac{\mu_{i}\delta^{2}}{2\lambda^{4}\left(n'+1\right)\left(n'+2\right)}+\frac{2\mu_{i}^{2}\delta}{\lambda^{4}\left(n'+2\right)\left(n'+4\right)}\right)
\end{eqnarray}
For any given set of fluid parameters, the only unknown that is needed to compute $I$ from Equation
\ref{eqICarTu3} is $\gamma_{w}$. Now, by definition, through the application of the main
rheological equation to the flow at the tube wall, we have

\begin{equation}
\mu_{w}\gamma_{w}=\tau_{w}
\end{equation}
that is

\begin{equation}
\left[\mu_{i}+\delta\left(1+\lambda^{2}\gamma_{w}^{2}\right)^{n'/2}\right]\gamma_{w}=\frac{R\,\Delta
p}{2L}
\end{equation}
From the last equation, $\gamma_{w}$ can be obtained numerically by a simple numerical solver,
based for example on a bisection method, and hence $I$ is computed from Equation \ref{eqICarTu3}.
The volumetric flow rate is then obtained from

\begin{equation}
Q=\frac{\pi R^{3}I}{\tau_{w}^{3}}
\end{equation}
which is fully analytical solution apart from computing the value of $\gamma_{w}$. However, since
obtaining $\gamma_{w}$ numerically can be easily achieved to any required level of accuracy, as it
only depends on very simple and reliable solution schemes like the bisection methods, this does not
affect the analytical nature of the solution and its accuracy.

\subsection{Cross}

For Cross fluids, the viscosity is given by \cite{OwensbookP2002}

\begin{equation}
\mu=\frac{\tau}{\gamma}=\mu_{i}+\frac{\mu_{o}-\mu_{i}}{1+\lambda^{m}\gamma^{m}}=\mu_{i}+\frac{\delta}{1+\lambda^{m}\gamma^{m}}
\end{equation}
where $\mu_{o}$ is the low-shear viscosity, $\mu_{i}$ is the high-shear viscosity, $\lambda$ is a
characteristic time constant, $m$ is an indicial parameter, and
$\delta=\left(\mu_{o}-\mu_{i}\right)$. Therefore

\begin{equation}
\tau=\gamma\left(\mu_{i}+\frac{\delta}{1+\lambda^{m}\gamma^{m}}\right)\label{tauEqCro}
\end{equation}
If we follow a similar procedure to that of Carreau by applying the WRMS method and labeling the
right hand side integral with $I$ we get

\begin{equation}
I=\int_{0}^{\tau_{w}}\gamma^{3}\left(\mu_{i}+\frac{\delta}{1+\lambda^{m}\gamma^{m}}\right)^{2}d\tau\label{eqICroTu}
\end{equation}
Now, from Equation \ref{tauEqCro} we have

\begin{equation}
d\tau=\left(\mu_{i}+\frac{\delta}{1+\lambda^{m}\gamma^{m}}-\frac{m\delta\lambda^{m}\gamma^{m}}{\left(1+\lambda^{m}\gamma^{m}\right)^{2}}\right)d\gamma\label{eqdtauCro}
\end{equation}
On substituting from Equation \ref{eqdtauCro} into Equation \ref{eqICroTu} and changing the
integration limits we get

\begin{equation}
I=\int_{0}^{\gamma_{w}}\gamma^{3}\left(\mu_{i}+\frac{\delta}{1+\lambda^{m}\gamma^{m}}\right)^{2}\left(\mu_{i}+\frac{\delta}{1+\lambda^{m}\gamma^{m}}-\frac{m\delta\lambda^{m}\gamma^{m}}{\left(1+\lambda^{m}\gamma^{m}\right)^{2}}\right)d\gamma\label{eqICroTu2}
\end{equation}
On solving this integral analytically and evaluating it at its two limits we obtain

\begin{eqnarray}\label{eqICroTu3}
I & = & \frac{\left\{ 2\delta^{3}\left[-m\left(2f^{2}+5f+3\right)+4g^{2}+2m^{2}\right]+12m\delta^{2}\mu_{i}g\left(m-g\right)+12m^{2}\delta\mu_{i}^{2}g^{2}+3m^{2}\mu_{i}^{3}g^{3}\right\} \gamma_{w}^{4}}{12m^{2}g^{3}} \nonumber \\
 & - & \frac{\left\{ \delta^{3}\left(m^{2}-6m+8\right)+3m\delta^{2}\mu_{i}\left(m-4\right)+3m^{2}\delta\mu_{i}^{2}\right\} {}_{2}F_{1}\left(1,\frac{4}{m};\frac{m+4}{m};-f\right)\gamma_{w}^{4}}{12m^{2}}
\end{eqnarray}
where

\begin{equation}
f=\lambda^{m}\gamma_{w}^{m}\,\,\,\,\,\,\,\,\,\,\,\,,\,\,\,\,\,\,\,\,\,\,\,\, g=1+f
\end{equation}
and $_{2}F_{1}$ is the hypergeometric function of the given argument with its real part being used
in this evaluation. As before, we have

\begin{equation}
\mu_{w}\gamma_{w}=\tau_{w}
\end{equation}
that is

\begin{equation}
\left(\mu_{i}+\frac{\delta}{1+\lambda^{m}\gamma_{w}^{m}}\right)\gamma_{w}=\frac{R\,\Delta p}{2L}
\end{equation}
From the last equation, $\gamma_{w}$ can be obtained numerically, e.g. by a bisection method, and
hence $I$ is evaluated. The volumetric flow rate is then computed from

\begin{equation}
Q=\frac{\pi R^{3}I}{\tau_{w}^{3}}
\end{equation}
which is fully analytical solution, as explained in the Carreau case.

\section{Slit Flow}\label{Slit}

In the following two subsections we apply the adapted WRMS method to derive analytical relations
for the flow of Carreau and Cross fluids in straight rigid uniform long thin slits.

\subsection{Carreau}

For Carreau fluids, the viscosity and shear stress are given by Equations \ref{eqCarVis} and
\ref{tauEqCar} respectively. Now, from the adapted WRMS method for slits we have

\begin{equation}
\frac{Q\tau_{w}^{2}}{2WB^{2}}=\int_{0}^{\tau_{w}}\gamma\tau d\tau\label{eqWRMSSl}
\end{equation}
If we label the integral on the right hand side of Equation \ref{eqWRMSSl} with $I$, substitute for
$\tau$ and $d\tau$ from Equations \ref{tauEqCar} and \ref{eqdtauCar} into $I$, and change the
integration limits we obtain

\begin{equation}
I=\int_{0}^{\gamma_{w}}\gamma^{2}\left[\mu_{i}+\delta\left(1+\lambda^{2}\gamma^{2}\right)^{n'/2}\right]\left[\mu_{i}+\delta\left(1+\lambda^{2}\gamma^{2}\right)^{n'/2}+n'\delta\lambda^{2}\gamma^{2}\left(1+\lambda^{2}\gamma^{2}\right)^{(n'-2)/2}\right]d\gamma\label{eqICarSl2}
\end{equation}
On solving this integral analytically and evaluating it at its two limits we obtain

\begin{eqnarray}\label{eqICarSl3}
I & = & \frac{n'\delta^{2}\gamma_{w}\left[_{2}F_{1}\left(\frac{1}{2},1-n';\frac{3}{2};-\lambda^{2}\gamma_{w}^{2}\right)-{}_{2}F_{1}\left(\frac{1}{2},-n';\frac{3}{2};-\lambda^{2}\gamma_{w}^{2}\right)\right]}{\lambda^{2}} \nonumber \\
 & + & \frac{\left(1+n'\right)\delta^{2}\gamma_{w}^{3}\,{}_{2}F_{1}\left(\frac{3}{2},-n';\frac{5}{2};-\lambda^{2}\gamma_{w}^{2}\right)}{3} \nonumber \\
 & + & \frac{n'\delta\mu_{i}\gamma_{w}\left[_{2}F_{1}\left(\frac{1}{2},1-\frac{n'}{2};\frac{3}{2};-\lambda^{2}\gamma_{w}^{2}\right)-{}_{2}F_{1}\left(\frac{1}{2},-\frac{n'}{2};\frac{3}{2};-\lambda^{2}\gamma_{w}^{2}\right)\right]}{\lambda^{2}} \nonumber \\
 & + & \frac{\left(2+n'\right)\delta\mu_{i}\gamma_{w}^{3}\,{}_{2}F_{1}\left(\frac{3}{2},-\frac{n'}{2};\frac{5}{2};-\lambda^{2}\gamma_{w}^{2}\right)+\mu_{i}^{2}\gamma_{w}^{3}}{3}
\end{eqnarray}
where $_{2}F_{1}$ is the hypergeometric function of the given arguments with the real part being
used in this evaluation. Now, from applying the rheological equation at the slit wall we have

\begin{equation}
\left[\mu_{i}+\delta\left(1+\lambda^{2}\gamma_{w}^{2}\right)^{n'/2}\right]\gamma_{w}=\frac{B\,\Delta p}{L}
\end{equation}
From the last equation, $\gamma_{w}$ can be obtained numerically by a simple numerical solver and
hence $I$ is computed. The volumetric flow rate is then obtained from

\begin{equation}
Q=\frac{2WB^{2}I}{\tau_{w}^{2}}
\end{equation}

\subsection{Cross}

If we follow a similar procedure to that of Carreau flow in slits by applying the adapted WRMS
method and labeling the right hand side integral with $I$ we get

\begin{equation}
I=\int_{0}^{\tau_{w}}\gamma^{2}\left(\mu_{i}+\frac{\delta}{1+\lambda^{m}\gamma^{m}}\right)d\tau\label{eqICroSl}
\end{equation}
where $d\tau$ is given by Equation \ref{eqdtauCro}. On substituting from Equation \ref{eqdtauCro}
into Equation \ref{eqICroSl} and changing the integration limits we get

\begin{equation}
I=\int_{0}^{\gamma_{w}}\gamma^{2}\left(\mu_{i}+\frac{\delta}{1+\lambda^{m}\gamma^{m}}\right)\left(\mu_{i}+\frac{\delta}{1+\lambda^{m}\gamma^{m}}-\frac{m\delta\lambda^{m}\gamma^{m}}{\left(1+\lambda^{m}\gamma^{m}\right)^{2}}\right)d\gamma\label{eqICroSl2}
\end{equation}
On solving this integral equation analytically and evaluating it at its two limits we obtain

\begin{equation}\label{eqICroSl3}
I=\frac{\left[3\delta^{2}\left(m-g\right)-\left\{
\delta^{2}\left(m-3\right)+2m\delta\mu_{i}\right\}
g^{2}\,_{2}F_{1}\left(1,\frac{3}{m};1+\frac{3}{m};-f\right)+6m\delta\mu_{i}g+2m\mu_{i}^{2}g^{2}\right]\gamma_{w}^{3}}{6mg^{2}}
\end{equation}
where

\begin{equation}
f=\lambda^{m}\gamma_{w}^{m}\,\,\,\,\,\,\,\,\,\,\,\,,\,\,\,\,\,\,\,\,\,\,\,\, g=1+f
\end{equation}
and $_{2}F_{1}$ is the hypergeometric function of the given argument with its real part being used
in this evaluation. As before, from applying the rheological equation at the wall we have

\begin{equation}
\left(\mu_{i}+\frac{\delta}{1+\lambda^{m}\gamma_{w}^{m}}\right)\gamma_{w}=\frac{B\,\Delta p}{L}
\end{equation}
From this equation, $\gamma_{w}$ can be obtained numerically and hence $I$ is computed. Finally,
the volumetric flow rate is obtained from

\begin{equation}
Q=\frac{2WB^{2}I}{\tau_{w}^{2}}
\end{equation}

\section{Validation}\label{Validation}

We validate the derived equations in the last two sections for the flow of Carreau and Cross fluids
through pipes and slits by two means. First, we validate the derived $I$ expressions (i.e.
Equations \ref{eqICarTu3}, \ref{eqICroTu3}, \ref{eqICarSl3} and \ref{eqICroSl3}), which are the
main source of potential errors in these derivations, by numerical integration of their
corresponding integrals (i.e. Equations \ref{eqICarTu2}, \ref{eqICroTu2}, \ref{eqICarSl2} and
\ref{eqICroSl2} respectively). On comparing the analytical solutions to those obtained from the
numerical integration of $I$, we obtained virtually identical results in all the investigated
cases. This validation eliminates the possibility of a formal error in the derivation of the
analytical expressions of $I$ but does not provide a proper validation to the basic WRMS method
since the numerical integration does not eliminate possible errors in the fundamental assumptions
and basic principles upon which the WRMS method is based and the derivation steps that lead to the
integrals $I$ and subsequently to $Q$. The second way of validation, which will be explained in the
following paragraphs, should provide this sort of validation as it is based on comparing the WRMS
solutions to solutions obtained from a totally different method, namely the variational method,
which was already validated.

In \cite{SochiVariational2013} a variational method based on applying the Euler-Lagrange
variational principle to find analytical and semi-analytical solutions for the flow of generalized
Newtonian fluids in straight rigid circular uniform pipes was proposed. There, the method was
applied to the Carreau and Cross fluids, among other types of fluid, where a mixed
analytical-numerical method was developed and used to find the flow of these two fluids in pipes.
Later on \cite{SochiSlitPaper2014, SochiVarNonNewt2014}, the variational method was further
validated by extending it to the slit geometry and to more types of non-Newtonian fluids, namely
Ree-Eyring and Casson. Elaborate details about all these issues, as well as other related issues,
can be found in the above references.

\begin{figure}
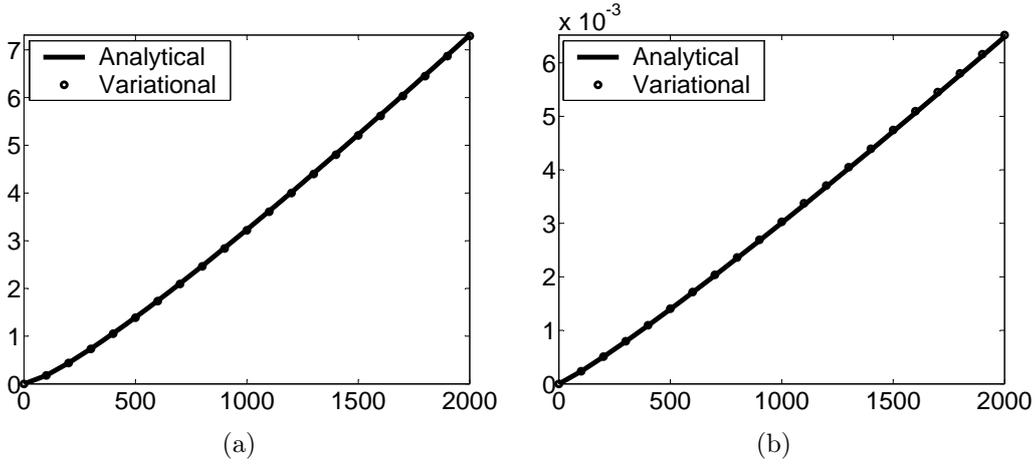

\centering %
\subfigure[]%
{\begin{minipage}[b]{0.5\textwidth} \CIF {g/CarreauTube1}
\end{minipage}}
\Hs
\subfigure[]%
{\begin{minipage}[b]{0.5\textwidth} \CIF {g/CarreauTube2}
\end{minipage}}
%
\caption{Comparing the WRMS analytical solutions to the variational solutions for the flow of
Carreau fluids in straight rigid circular uniform tubes with
(a) $n=0.65$, $\mu_o=0.1$~Pa.s, $\mu_{i}=0.005$~Pa.s, $\lambda=1.5$~s, $L=0.85$~m and $R=0.09$~m;
and (b) $n=0.9$, $\mu_o=0.08$~Pa.s, $\mu_{i}=0.001$~Pa.s, $\lambda=2.0$~s, $L=0.5$~m and
$R=0.02$~m.
In both sub-figures, the vertical axis represents the volumetric flow rate, $Q$, in m$^3$.s$^{-1}$
while the horizontal axis represents the pressure drop, $\Delta p$, in Pa. The average percentage
relative difference between the analytical and variational solutions for these cases are about
0.37\% and 0.60\% respectively. \label{CarreauTubeFig}}
\end{figure}

\begin{figure}
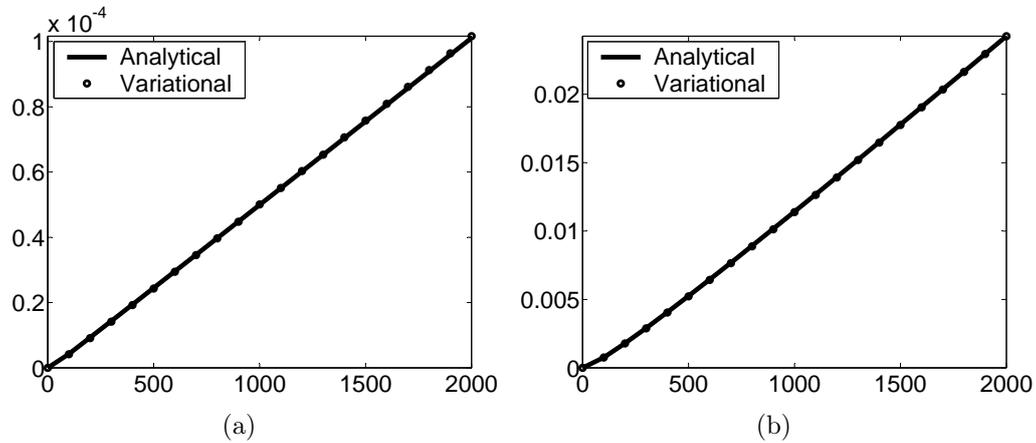

\centering %
\subfigure[]%
{\begin{minipage}[b]{0.5\textwidth} \CIF {g/CrossTube1}
\end{minipage}}
\Hs
\subfigure[]%
{\begin{minipage}[b]{0.5\textwidth} \CIF {g/CrossTube2}
\end{minipage}}
%
\caption{Comparing the WRMS analytical solutions to the variational solutions for the flow of Cross
fluids in straight rigid circular uniform tubes with
(a) $m=0.83$, $\mu_o=0.22$~Pa.s, $\mu_{i}=0.033$~Pa.s, $\lambda=6.65$~s, $L=0.95$~m and
$R=0.008$~m;
and (b) $m=0.5$, $\mu_o=0.15$~Pa.s, $\mu_{i}=0.009$~Pa.s, $\lambda=7.9$~s, $L=1.65$~m and
$R=0.027$~m.
In both sub-figures, the vertical axis represents the volumetric flow rate, $Q$, in m$^3$.s$^{-1}$
while the horizontal axis represents the pressure drop, $\Delta p$, in Pa. The average percentage
relative difference between the analytical and variational solutions for these cases are about
0.55\% and 0.32\% respectively. \label{CrossTubeFig}}
\end{figure}

\begin{figure}
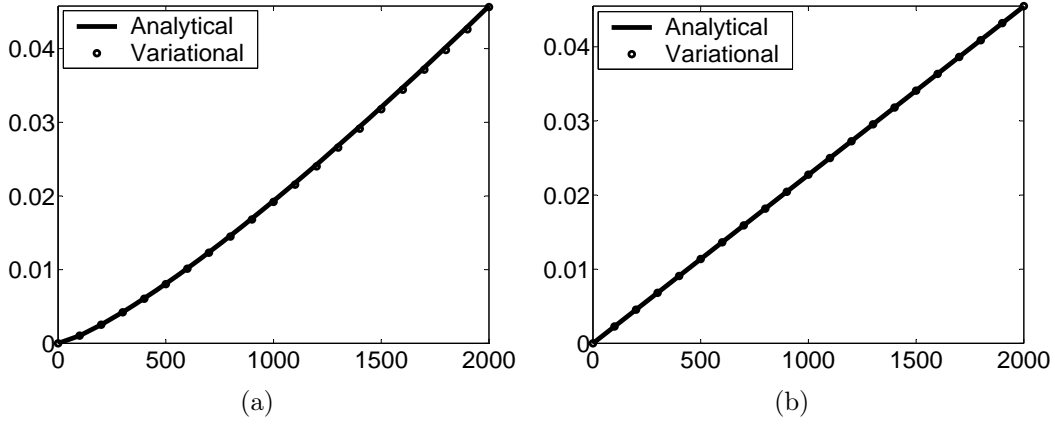

\centering %
\subfigure[]%
{\begin{minipage}[b]{0.5\textwidth} \CIF {g/CarreauSlit1}
\end{minipage}}
\Hs
\subfigure[]%
{\begin{minipage}[b]{0.5\textwidth} \CIF {g/CarreauSlit2}
\end{minipage}}
%
\caption{Comparing the WRMS analytical solutions to the variational solutions for the flow of
Carreau fluids in straight rigid uniform long thin slits with
(a) $n=0.75$, $\mu_o=0.17$~Pa.s, $\mu_{i}=0.009$~Pa.s, $\lambda=2.5$~s, $L=1.3$~m, $W=1.0$~m and
$B=0.012$~m;
and (b) $n=1.0$, $\mu_o=0.09$~Pa.s, $\mu_{i}=0.012$~Pa.s, $\lambda=3.0$~s, $L=1.1$~m, $W=1.0$~m and
$B=0.015$~m.
In both sub-figures, the vertical axis represents the volumetric flow rate, $Q$, in m$^3$.s$^{-1}$
while the horizontal axis represents the pressure drop, $\Delta p$, in Pa. The average percentage
relative difference between the analytical and variational solutions for these cases are about
0.70\% and 0.06\% respectively. \label{CarreauSlitFig}}
\end{figure}

\begin{figure}
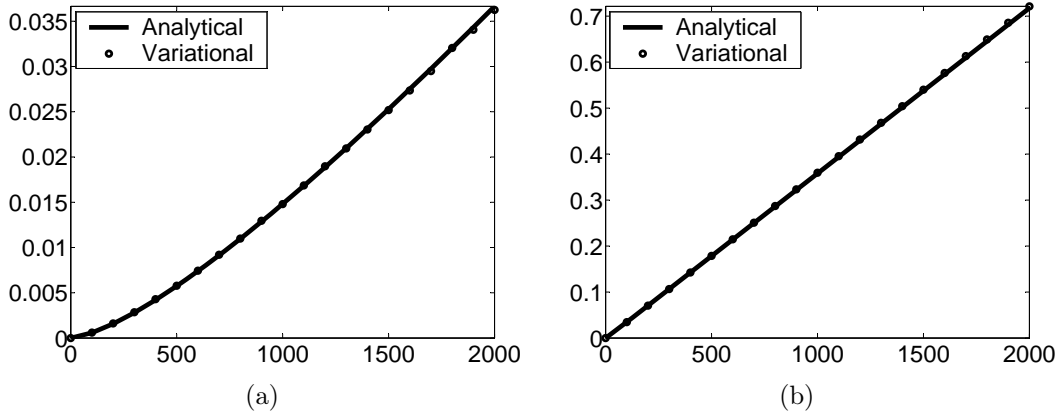

\centering %
\subfigure[]%
{\begin{minipage}[b]{0.5\textwidth} \CIF {g/CrossSlit1}
\end{minipage}}
\Hs
\subfigure[]%
{\begin{minipage}[b]{0.5\textwidth} \CIF {g/CrossSlit2}
\end{minipage}}
%
\caption{Comparing the WRMS analytical solutions to the variational solutions for the flow of Cross
fluids in straight rigid uniform long thin slits with
(a) $m=0.45$, $\mu_o=0.08$~Pa.s, $\mu_{i}=0.003$~Pa.s, $\lambda=0.75$~s, $L=0.75$~m, $W=1.0$~m and
$B=0.005$~m;
and (b) $m=0.75$, $\mu_o=0.03$~Pa.s, $\mu_{i}=0.005$~Pa.s, $\lambda=5.0$~s, $L=1.25$~m, $W=1.0$~m
and $B=0.015$~m.
In both sub-figures, the vertical axis represents the volumetric flow rate, $Q$, in m$^3$.s$^{-1}$
while the horizontal axis represents the pressure drop, $\Delta p$, in Pa. The average percentage
relative difference between the analytical and variational solutions for these cases are about
0.89\% and 0.40\% respectively. \label{CrossSlitFig}}
\end{figure}

Extensive comparisons between the variational and the WRMS analytical solutions have been carried
out as part of the current investigation to validate the analytical solutions on one hand and to
add more support to the variational method on the other. In all the investigated cases, which vary
over a wide range of fluid and conduit parameters, very good agreement was obtained between the
variational and the WRMS analytical solutions. In Figures \ref{CarreauTubeFig}--\ref{CrossSlitFig}
we present a representative sample of the investigated cases where we compare the WRMS analytical
solutions with the variational solutions for the flow of Carreau and Cross fluids through pipe and
slit geometries. As seen in these figures, the agreement between the two methods is excellent which
is typical of all the investigated cases. The main source of error and departure between the two
methods is the heavy use of numerical bisection solvers and numerical integration in the
implementation of the variational method. The presence of the hypergeometric function in both the
WRMS and variational solutions is another source of error since this complicated function may not
converge satisfactorily in some cases when evaluated numerically, and hence becomes problematic and
a major source of error. It is worth mentioning that the solutions obtained from the numerical
integration of $I$ integrals are not shown on Figures \ref{CarreauTubeFig}--\ref{CrossSlitFig}
because they are virtually identical to the analytical solutions.

\clearpage

\clearpage

\clearpage
\section{Conclusions} \label{Conclusions}

In this paper, a method based on the application of the Weissenberg-Rabinowitsch-Mooney-Schofield
(WRMS) equation for the flow of generalized Newtonian fluids in straight rigid circular uniform
pipes and its extension to straight rigid long thin uniform slits has been developed and used to
find analytical flow solutions for Carreau and Cross fluids which do not have known analytical
solutions in those geometries. The main analytical expressions were verified by numerical
integration to rule out potential errors in the formality of derivation. The analytical solutions
were then thoroughly compared to the solutions of another method, namely the variational approach
which is based on the application of the Euler-Lagrange principle and hence it is totally different
and independent from the WRMS method. Excellent agreement was obtained in all the investigated
cases.

The first thing that has been achieved in this investigation is obtaining fully analytical
solutions for the flow of Carreau and Cross fluids in the above mentioned geometries. These
solutions provide a better alternative to the use of numerical techniques, which are currently the
only available means, since the analytical solutions are easier to obtain, less prone to error and
highly accurate. The second is that these analytical solutions provide more support to the
previously proposed variational method for obtaining solutions for the flow of generalized
Newtonian fluids in confined geometries. In fact this agreement should serve as mutual validation
since these methods are totally independent and are based on very different theoretical and
mathematical infrastructures and hence they lend support to each other.

It should be remarked that the derivation of the pipe and slit analytical relations for the Carreau
model can be easily extended to the more general Carreau-Yasuda model by replacing `2' with `$a$'
in the exponents of the Carreau constitutive relation. The WRMS method may also be applied to other
non-Newtonian fluid models which do not have analytical solutions. However, in some cases the
definite integrals, $I$, may require numerical evaluation by using one of the numerical integration
methods such as quadratures, Simpson, trapezium or midpoint integration rules if fully analytical
solutions are very difficult or impossible to obtain. Such semi-analytical solutions will not only
be much easier to obtain and less prone to errors than their numerical counterparts, but can also
be as accurate as any potential analytical solutions for all practical purposes due to the
availability of very reliable numerical integration schemes that can be employed. These schemes are
easier to implement and more reliable than the alternative numerical techniques like those based on
discretization schemes.

\clearpage
\section{Nomenclature}

\begin{supertabular}{ll}
$\gamma$                &   shear rate (s$^{-1}$) \\
$\gamma_w$              &   shear rate at conduit wall (s$^{-1}$) \\
$\delta$                &   $\mu_{o}-\mu_{i}$ (Pa.s) \\
$\lambda$               &   characteristic time constant (s) \\
$\mu$                   &   dynamic shear viscosity of generalized Newtonian fluid (Pa.s) \\
$\mu_{i}$               &   high-shear viscosity (Pa.s) \\
$\mu_{o}$               &   low-shear viscosity (Pa.s) \\
$\tau$                  &   shear stress (Pa) \\
$\tau_B$                &   shear stress at slit wall (Pa) \\
$\tau_R$                &   shear stress at tube wall (Pa) \\
$\tau_w$                &   shear stress at conduit wall (Pa) \\
\\
$A_{\parallel}$         &   area of conduit internal surface (m$^2$) \\
$B$                     &   slit half height (m) \\
$f$                     &   $\lambda^{m}\gamma_{w}^{m}$ \\
$F_{\perp}$             &   force normal to conduit cross section (N) \\
$_{2}F_{1}$             &   hypergeometric function \\
$g$                     &   $1+f$ \\
$I$                     &   definite integral expression (Pa$^3$.s$^{-1}$ for pipe and Pa$^2$.s$^{-1}$ for slit) \\
$L$                     &   length of conduit (m) \\
$m$                     &   indicial parameter in Cross model \\
$n$                     &   flow behavior index in Carreau model \\
$n'$                    &   $n-1$ \\
$\Delta p$              &   pressure drop (Pa) \\
$Q$                     &   volumetric flow rate (m$^{3}$.s$^{-1}$) \\
$r$                     &   radius (m) \\
$R$                     &   tube radius (m) \\
$v$                     &   fluid velocity in the flow direction (m.s$^{-1}$) \\
$W$                     &   slit width (m) \\
$z$                     &   coordinate of slit smallest dimension (m) \\
\end{supertabular}

\clearpage
\phantomsection \addcontentsline{toc}{section}{References} %
\bibliographystyle{unsrt}

\end{document}